\documentstyle[psfig]{ioplppt}      

\def\bbbq{{\mathchoice {\setbox0=\hbox{$\displaystyle\rm Q$}\hbox{\raise
0.15\ht0\hbox to0pt{\kern0.4\wd0\vrule height0.8\ht0\hss}\box0}}
{\setbox0=\hbox{$\textstyle\rm Q$}\hbox{\raise
0.15\ht0\hbox to0pt{\kern0.4\wd0\vrule height0.8\ht0\hss}\box0}}
{\setbox0=\hbox{$\scriptstyle\rm Q$}\hbox{\raise
0.15\ht0\hbox to0pt{\kern0.4\wd0\vrule height0.7\ht0\hss}\box0}}
{\setbox0=\hbox{$\scriptscriptstyle\rm Q$}\hbox{\raise
0.15\ht0\hbox to0pt{\kern0.4\wd0\vrule height0.7\ht0\hss}\box0}}}}
\def\bbbz{{\mathchoice {\hbox{$\sf\textstyle Z\kern-0.4em Z$}}
{\hbox{$\sf\textstyle Z\kern-0.4em Z$}}
{\hbox{$\sf\scriptstyle Z\kern-0.3em Z$}}
{\hbox{$\sf\scriptscriptstyle Z\kern-0.2em Z$}}}}

\begin{document}

\jl{1}

\title{Trace identities and their semiclassical implications}
\author{ Uzy Smilansky \footnote{Permanent address: Department of
Physics of Complex Systems,  The Weizmann Institute of Science, Rehovot 76100,
Israel}\\[5mm]}
\address {Fachbereich Physik, Philipps-Universitaet Marburg,
D-35032 Marburg, Germany\\[2mm]}
\date{\today}

\begin{abstract}
  The compatibility of the semiclassical quantization of area-preserving maps
with some exact identities which follow from the unitarity of the quantum
evolution
operator is discussed. The quantum identities involve relations between
traces  of
powers of the  evolution operator. For classically {\it integrable} maps, the
semiclassical approximation is shown to be compatible with the trace
identities. This
is  done by the identification of stationary phase manifolds which give the
main
contributions to the result. The same technique is not applicable for {\it
chaotic}
maps, and the compatibility of the semiclassical theory in this case remains
unsettled. The compatibility of the semiclassical
quantization with the trace identities  demonstrates the crucial importance of
non-diagonal contributions.
\end{abstract}
\section{\bf Introduction}
 \label {introduction}

 This paper focuses on quantum maps which are represented by unitary evolution
operators  on a Hilbert space of a finite dimension. The quantum map
propagates any
initial state in the Hilbert space by
\begin{equation}
\psi_{t+1} = U \psi_t  = U^{t+1}\psi_0 \ .
\end{equation}
 It is assumed that there exists an  underlying  area
preserving classical map which acts on a compact phase space ${\cal M}$,
which can  be
considered as the counterpart of the quantum map within the semiclassical
approximation.  That is,

\noindent {\it (i)} The semiclassical quantization of the classical map
(see section
 \ref {sec:sclquant}) provides an approximation to the exact quantum map. The
semiclassical approximant is unitary $ U_{\rm scl} ( U_{\rm
scl})^{\dagger}=I$ and
satisfies the composition rule
$U^{t+s}_{\rm scl}=U^t_{\rm scl} U^s_{\rm scl}$
within the accuracy margin of the semiclassical approximation.

\noindent  {\it (ii)} The dimension of the Hilbert space, $M$, is related
to the
classical phase space volume $|{\cal M} |$ by
\begin{equation}
M = \left [ {|{\cal M}|\over (2\pi \hbar)^f } \right ]
\label{eq: dimension}
\end{equation}
where $\left [ \cdot \right ]$ stands for the integer part, and $f$ is the
number of
classical  freedoms.

 Consider a quantum map $U$ and introduce the notation  $t_n \equiv {\rm
tr} U^n$. The
fact that $U$ is unitary  imposes various relations amongst the $t_n$ which
should be
satisfied identically. The  purpose of this paper is to study to what
extent the
semiclassical approximation is compatible with  a certain class of
identities, which,
to the best of my knowledge, was not examined in this context till now. To
give an
 idea, the most simple version of the identities to be considered is
\begin {eqnarray}
\lim_{\epsilon \rightarrow 0} \ \epsilon \ \sum _{n=n_0}^{\infty}
t_n^{*}t_{n+\nu}
{\rm e}^{-n\epsilon}   = t_{\nu} \ ,
\nonumber
\end{eqnarray}
for arbitrary integers $n_0$ and $\nu$.  These identities will be proved and
discussed in detail in  section  \ref {sec:identities} . However, before
doing this,
it is instructive to review a  few other identities, which involve the
$t_n$,  and
which were  used in past investigations of the semiclassical approximation
\cite{{Bogomol},{DoronUS},{USLeshouches},{haake}}.

 The first class of identities can be derived by studying the properties of the
characteristic polynomial
\begin{equation}
p_{_U}(z) \equiv \det (I-zU) =\sum_{m=0}^M a_m z^m \ .
\label {polynom}
\end{equation}
Since $p_{_U}(U^{\dagger}) = 0 $,
\begin{equation}
a_0 = 1 = - \sum_{m=1}^M a_m(U^{\dagger})^m  \ .
\end {equation}
Multiplying  by $U^n$  ($n>M$) and taking the trace  one gets
\begin{equation}
t_n  = - \sum_{m=1}^M a_m  t_{n-m} \ .
\label {sumtm}
\end {equation}
By successive  applications of the above relation, all the traces  $t_n $ with
$n>M$ can be expressed in terms  of the traces of the $M$ lowest powers.

An important consequence of the unitarity of $U$ is the {\it inversive
symmetry} of
the coefficients $a_m$,
 \begin{equation}
 a_m = e^{i\Theta}a^\ast_{M-m} \  ,
\label {inversive}
\end{equation}
 where $\det (-U) \equiv e^{i \Theta} $. One can utilize the inversive
symmetry to
 obtain identities between the $t_n$ by invoking
 Newton's identities. They  relate the traces $t_n $
 and the coefficients of the secular polynomial $a_m$:
 \begin{equation}
 a_m = -{1\over m} \left(t_m +\sum_{k=1}^{m-1}a_k t_{m-k} \right ) \ .
\label {Newton}
\end{equation}
  Since $a_m =0$ for $m > M$,  the $t_n$ for all $n > M$ depend linearly on the
lower $m \le M$ traces, which is consistent with our previous observation.
Successive iterations  yield explicit expressions for the coefficients $a_m$
in terms of the $t_n$, and one can substitute them in  (\ref {sumtm}) or in
(\ref {inversive}).

The significance of such relations in the semiclassical context is due to
the fact
that the $t_n$ are expressed semiclassically as sums over $n$-periodic
orbits of the
classical map. Thus, the compatibility with the exact identities implies that
there exist  identities relating sums over periodic orbits of {\it
different} periods
which are satisfied within the margins of the semiclassical accuracy.
The resulting identities  between the $t_n$ are getting complicated as
$n$ and $M$ increase, and therefore their compatibility with the semiclassical
approximation was seldom tested  \cite {EckhardtHaake}.

 Another set of identities which will be shown to be closely related to the
present
work were introduced by  M.V. Berry in  \cite {BerryA400}. He considered
the spectral
density of a quantum hamiltonian
\begin{equation}
d(E) = \sum_{n=1}^{\infty} \delta (E-E_n) \ ,
\end{equation}
and its smooth approximant
\begin{equation}
d_{\epsilon}(E) = \sum_{n=1}^{\infty} \delta_{\epsilon} (E-E_n) \ \ {\rm
with} \ \
\delta_{\epsilon}(x) =  {1\over \pi} {\epsilon\over \epsilon^2 +x^2} \ .
\end{equation}
Assuming that the spectrum has no degeneracies ($E_n \ne E_m$  when $n\ne
m$), one
finds,
\begin{equation}
2\pi \lim _{\epsilon \rightarrow 0}\  \epsilon \  d_{\epsilon}^2 (E) = d(E) \ .
\label {eq:bootstrap}
\end{equation}

 Substituting the semiclassical trace formula in both sides of  (\ref
{eq:bootstrap})
one sees that the left hand side is quadratic while the right hand side is
linear in
the periodic orbit amplitudes. Integrating (\ref {eq:bootstrap}) over a
sufficiently
large energy domain,  the contributions of long orbits to the right hand
side can be
made arbitrarily small, while their contribution to the left hand side will
not be
smoothed out. This observation led  Berry to conclude  that the long
periodic orbits
must contain information about the short periodic orbits if the semiclassical
approximation is compatible with (\ref {eq:bootstrap}). This information is
stored
as correlations  between actions of periodic orbits, because pairs
of {\it distinct} periodic orbits combine together to give an amplitude of
order
$\epsilon ^{-1}$ in $ d_{\epsilon}^2 (E)$, which, upon multiplying by
$\epsilon$
reproduce the periodic orbit contributions to the oscillatory parts of
$d(E)$. This is a highly ``non-diagonal" effect, which needs very
special correlations between the actions to get the correct result. This
observation
shows that the use of identities of this type comes naturally  in the
context of the
study of classical action correlations and their effect on the statistics
of the
quantum spectra  \cite{Argaman,Dittes,doroncohen}.

J. Keating \cite {Keat91} (see also \cite {Connors}) generalized (\ref
{eq:bootstrap})
and used it in his studies of the spectrum of the Riemann zeros. The
``non-diagonal"
correlations which are necessary to prove the identities are introduced by
using
 the Hardy-Littlewood conjecture on the correlations between primes.

 E. Bogomolny \cite {Bogo99} tested (\ref {eq:bootstrap}) for the spectrum of a
rectangular billiard with periodic boundary conditions. By considering
carefully the
stationary phase manifolds in the sums over periodic orbits which arise
from the left
hand side, he was able to perform the summations and to demonstrate the
compatibility
of the semiclassical approximation with (\ref {eq:bootstrap}). The same
methods were
later used to analyze  integrable systems in general. The present work was
inspired by a seminar given by Professor Bogomolny, and it can be
considered as an
extension of Bogomolny's ideas to quantum maps and their semiclassical
approximation.

  This manuscript is arranged in the following way.
The trace identities which will be the main tool in the present analysis
will be
derived in the next section. The  semiclassical quantization of area
preserving maps
will be reviewed in section   \ref {sec:sclquant} and the compatibility of the
semiclassical approximation will be demonstrated in  section  \ref
{sec:compat} for
integrable maps. The close relation between the compatibility problem and the
correlations between actions of periodic orbits will be addressed at the
end of this
section.  The difficulties encountered in attempting to use the same
methods to study
the compatibility in the case of  hyperbolic maps are discussed  as well.
By applying
the trace identities to certain quantum graphs, it is possible to derive
combinatorial
identities which involve Krawtchouk polynomials \cite {kp1,kp2}. These
polynomial are important building blocks in the theory of error correcting
codes \cite
{cohen}, and the  identities might be of use in this  branch of
mathematics. Since
this application is not directly connected to the issue of the semiclassical
compatibility, it will be presented in the appendix.

\section {\bf Trace identities}
\label {sec:identities}

 Consider a unitary matrix $U$ of dimension $M$. Its spectrum consists of
$M$ points
on the unit circle $\{ {\rm e}^{i\theta_m}, \  m=1,\cdots ,M \} $, where the
eigenphases are real and are assumed to be distinct. Recalling the notation
$t_n\equiv
{\rm tr} U^n$, the following identities hold for arbitrary integers $n_0$
and $\nu$ :
\begin {equation}
\lim_{\epsilon \rightarrow 0} \ \epsilon \ \sum _{n=n_0}^{\infty}
t_n^{*}t_{n+\nu}
{\rm e}^{-n\epsilon}   = t_{\nu} \ .
\label{eq:identity1}
\end{equation}
 To prove these relations  one substitutes
$t_n=\sum_{m=1}^M{\rm e}^{i n\theta_m}$ and after summing the geometric
series, one
uses
\begin {equation}
\lim_{\epsilon \rightarrow 0} \ {\epsilon \over 1-{\rm e}^{i
(\theta_m-\theta_{m'})
-\epsilon } } =
 \left \{
\begin{array}{l}
  1 \ \ \  {\rm for}\ \ \ (\theta_m-\theta_{m'}) =0 \nonumber \\
  0 \ \ \  {\rm for}\ \ \ (\theta_m-\theta_{m'}) \ne 0
\end{array}
\right \} = \delta_{m,m'}  \ .
\label{eq:kronnecker}
\end{equation}
 The condition that there  are no degeneracies in the  spectrum of $U$ is
used to
justify the rightmost  equality in (\ref {eq:kronnecker}) .

  A few points are worth noticing:

\noindent - The $\epsilon \rightarrow 0$ limit of the sum weighted by
${\rm e}^{-n\epsilon}$  can be interpreted as a time average
\begin {equation}
\lim_{\epsilon \rightarrow 0} \ \epsilon \ \sum _{n=n_0}^{\infty}
(\cdot)_n\  {\rm
e}^{-n\epsilon}   = \lim _{N\rightarrow \infty} {1\over N}\sum_{n=n_0}^{n_0
+N}
(\cdot)_n
\ .
\label{eq:timeaverage}
\end{equation}

\noindent -  For $\nu=0$, and using  $t_0 = M$,  one gets
\begin {equation}
\lim_{\epsilon \rightarrow 0} \ \epsilon \ \sum _{n=0}^{\infty}{1\over M}
|t_n|^2
{\rm e}^{-n\epsilon}   = 1 \ .
\label{eq:identity0}
\end{equation}
 Thus, the time average of  ${1\over M} |t_n|^2$ approaches $1$, a result
familiar
from the study of the spectral form-factor for unitary matrices  \cite {Dyson}.

\noindent - The spectral density of $U$ on the unit circle can be written as
\begin {equation}
d(\theta) =\sum_{l=1}^M \delta (\theta-\theta_l) = \lim_{\epsilon\rightarrow 0}
d_{\epsilon}(\theta)   \ ; \   d_{\epsilon}(\theta) \equiv  {1\over 2\pi}
\sum_{m=-\infty}^{\infty} t_m {\rm e }^{-i m \theta-\epsilon |m|} \ .
\end{equation}
 Using  (\ref {eq:identity1}), one can  easily derive
\begin{equation}
2\pi \lim _{\epsilon \rightarrow 0}\  \epsilon \  d_{\epsilon}^2 (\theta) =
d(\theta)
\ ,
\label {eq:bootstraptheta}
\end{equation}
which is the analogue of (\ref {eq:bootstrap}) for the spectrum of unitary
operators.
\bigskip

 The following identities which involve products of more than two traces
can be proven
with the help of (\ref {eq:kronnecker})
\begin {eqnarray}
\label{eq:identity2}
 t_{\nu} & =&  \lim_{\epsilon_1 \rightarrow 0} \cdots \lim_{\epsilon_k
\rightarrow 0}   \epsilon_1 \cdots \epsilon_k\\
& &   \sum_{n_1= n_{1_0}}^{\infty}\cdots \sum_{n_k=n_{k_0}}^{\infty}
 t^*_{n_1}\cdots t^*_{n_k}  \   t_{(n_1 +\cdots +n_k)+\nu}\ \exp (-
\sum_{i=1}^k n_i\epsilon_i) \ , \nonumber
\end{eqnarray}
where the $n_{i_0}$ are arbitrary integers. The identities (\ref
{eq:identity2}) can be
also written as
\begin {eqnarray}
\label{eq:identity3}
 t_{\nu_1+ \cdots  +\nu_k}  &=&  \lim_{\epsilon_1 \rightarrow 0} \cdots
\lim_{\epsilon_k
\rightarrow 0}   \epsilon_1 \cdots \epsilon_k \\
&&
\sum_{n_1= n_{1_0}}^{\infty}\cdots \sum_{n_k=n_{k_0}}^{\infty}
t^*_{n_1+ \cdots + n_k}   t_{n_1+\nu_1}\cdots t_{n_k+\nu_k}\ \exp (-
\sum_{i=1}^k
n_i\epsilon_i)  . \nonumber
\end{eqnarray}
 The equivalence of (\ref {eq:identity2}) and (\ref {eq:identity3}) can be
shown
by taking the complex conjugate of (\ref {eq:identity3}), denoting $\nu_1+
\cdots
+\nu_k = -\nu$ and shifting the summation variables $n_i$ by  $\nu _i$. The
shifts
do  not affect the result since they can be absorbed in the arbitrary
$\{n_{i_0}\}$.

 The compatibility of the semiclassical approximation for $t_n$ with the
identities
(\ref {eq:identity1},\ref {eq:identity2}) will be  investigated
in section  \ref {sec:compat}, after the semiclassical approximation for
$t_n$ will be
reviewed in the following section.

\section  {\bf Semiclassical quantization of maps}
\label {sec:sclquant}

  The quantum unitary operator $U$ to be investigated,
is assumed to be the analogue of an area preserving map $\cal{F}$ acting
on a finite phase space domain $\cal{M}$ with area $|\cal{M}|$. (For the
sake of
simplicity the maps to be considered act on manifolds with $f=1$, and will
be assumed
to have the twist property.  The semiclassical treatment can be
extended to the general case.)   The phase space coordinates are denoted by
$\gamma =
(q,p)$  and $\gamma$ is mapped to $\bar \gamma  \equiv \cal{F} (\gamma)$.
Area--preserving  maps can be derived from a generating function (action)
$\Phi(q,\bar q )$
\begin{equation}
p= - {\partial\Phi(q,\bar q ) \over \partial q} \ \ \ ; \ \ \ \bar p=
{\partial\Phi(q,\bar q) \over \partial \bar q}.
\label{eq:classmap}
\end{equation}
The explicit mapping function $\bar \gamma = \cal {F} (\gamma) $ is
obtained by solving the implicit relations (\ref {eq:classmap}). The twist
condition
ensures that the implicit  equations (\ref {eq:classmap}) have a unique
solution.

In the case the map is  integrable, let $I$ be the invariant momentum (action
variable) under the action of the map and $\phi$ the canonically conjugate
angle
variable. The domain of the map is the annulus $I\in [I_{min},I_{max}] ,
\phi \in
[0,2\pi)$, and $|{\cal M}|=2\pi (I_{max}-I_{min})$ . In this
representation,  the
generating function must take the form
$\Phi(\phi,\bar \phi )=\Phi(\bar \phi-\phi)$. The explicit  map is
\begin{equation}
\bar I =I \ \ \ ; \ \ \ \Delta \phi = \bar \phi-\phi = f(I)
\label{integrablemap}
\end{equation}
Here, $f(I)$ (the angular velocity) is the  inverse of the
 generating relation $I = \Phi '(\Delta \phi)$, which gives $\Delta \phi$
in terms of
$I$. The twist condition is fulfilled if $ \Phi ''(\Delta \phi) \ne 0 $.

A classical trajectory is obtained by applying the map to an arbitrary initial
point in $\cal {M}$, and the corresponding action is accumulated along the
trajectory.

The semiclassical quantization of  ${\cal F} $ in
the $q$ representation is \cite {Miller,USLeshouches}:
\begin{equation}
\langle q|U|\bar q\rangle_{\rm scl} =\left ({1 \over2\pi \hbar i}\right )
^{1\over 2}
\left [{ \partial^2 \Phi(q,\bar q ) \over \partial q
\partial \bar q }\right ] ^{1\over 2}
{\rm e} ^{ {i\over \hbar} \Phi(q,\bar q ) }
\label {sclsmatrix}
\end{equation}
It can be shown to preserve the composition property and to be unitary
within the
semiclassical approximation.

 The semiclassical approximation for $t_n$ involves the periodic manifolds
of the
classical map.  For hyperbolic maps \cite {tabor,USLeshouches},
\begin{equation}
\left [ t_n \right ]_{\rm scl} = \sum_{p\in {\cal P}_n} { n_p e^{i r
              (\Phi_p /\hbar -\nu_p {\pi\over 2})} \over |\det
              (I-T^r_p)|^{{1\over 2}} } \ .
\label{trsemicl}
\end{equation}
 The semiclassical approximation for $t_n$ involves the set of $n$-periodic
orbits
 ${\cal P}_n$  which are repetitions of  primitive periodic orbits of ${\cal
F}$, with  periods $n_p$ which are divisors of $n$, so that $ n=n_p r$.
 The monodromy matrix is denoted by  $T_p$. Each periodic orbit contribution is
endowed with a phase which is the action summed along the periodic orbit,
\begin{equation}
 \Phi _p = \sum_{j=1}^{n_p} \Phi(q_j,q_{j+1}) \ \ \ ({\rm with} \ \ \ \ \
q_{n_p +1} = q_1 ),
\end {equation}
and of the Maslov contribution  $-\nu_p \pi/ 2$.

 For integrable maps, the action-angle variables $(I,\phi)$ will be used,
where $I$ is the classical invariant. In the quantum picture,
$I$ is quantized to integer multiples of $\hbar$ so that  $I_j = j\hbar$
and $1\le
j\le M$. The matrix $U$ is diagonal
in the action  representation. The semiclassical approximation for the
eigenphases can
be carried out directly,
\begin{equation}
\left [\theta_j\right ]_{\rm scl} =  {1\over \hbar} \left [ \Phi(f(I_j)) -
I_j f(I_j)
\right ]
\ .
\label{eq:thetascl}
\end{equation}
where  $f(I)$  is the angular frequency  (\ref {integrablemap}). Thus,
\begin {equation}
[ t_n]_{\rm scl} = \sum_{j=1}^M {\rm e}^{in[\theta_j]_{\rm scl}} \ ,
\label {eq:traceform}
\end{equation}
This semiclassical expression is not of the desired form, because it does
not express
$t_n$ in terms of the periodic orbits. However, performing the $j$ sum
using Poisson
summation, one gets,
\begin{equation}
\left [ t_n\right ]_{\rm scl} = \left ({2\pi \over n \hbar}\right )^{1\over2}
{\rm e}^ {- i( n+{1\over 2}) {\pi \over 2}}
\sum _{m=1}^n
\left [ \Phi''(\Delta\phi=2\pi
{m\over n}) \right]^{1\over2} {\rm e}^ {{ i\over \hbar}
n\Phi(\Delta\phi=2\pi {m\over
n}) } \ .
\label{eq:trinteg}
\end{equation}
Now, $t_n$ is expressed as a  sum  over the periodic manifolds of period
$n$ and winding number $m$. They occur at values of $I$ for which the
angular frequency
is rational $f(I_{n,m})  = 2\pi {m\over n}$.

  The expressions for $t_n$ in terms of periodic manifolds in the cases of
classically
integrable and classically chaotic maps are the necessary building blocks
for the
discussions which follows.

\section {\bf Compatibility of the semiclassical approximation}
\label {sec:compat}

 The compatibility of the semiclassical approximation for the $t_n$ with
the trace
identities will be now shown for {\it integrable} maps.
Turning first to (\ref {eq:identity1}),  the semiclassical expression
(\ref {eq:trinteg}) for $t_n$  is substituted in the right hand side of (\ref
{eq:identity1}) resulting in
\begin{eqnarray}
\label {eq:firstsum}
&&\epsilon \sum _{n=n_ 0}^{\infty} \left [t_n^{*}t_{n+\nu}\right ]_{\rm
scl} {\rm
e}^{-n\epsilon}\ = \\ &=& \  \epsilon\ {\rm e}^{-i\nu {\pi \over 2}} {2\pi
\over
\hbar}\sum _{n=n_ 0}^{\infty}  {{\rm e}^{-n\epsilon}\over (n(n+\nu))^{1\over2}}
\times \nonumber \\
&\times & \sum_{m=1}^n \sum_{m'=1}^{n+\nu}
\left ( \Phi''(2\pi {m\over n}) \Phi''(2\pi {m'\over n+\nu }) \right
)^{1\over2}
{\rm e}^{{i\over \hbar}
  [(n+\nu)\Phi(2\pi {m'\over n+\nu} )-n \Phi (2\pi {m \over n  }) ]}
\nonumber
\end{eqnarray}

  The sum above runs over the periodic manifolds (tori) of the map. It is
important
to respect the integer character of $n, m, m'$, since only for integer
values,  the
classical orbits are periodic. If we turn these sums to integrals by e.g.
Poisson
summation, we would loose this feature. To get a finite contribution for (\ref
{eq:firstsum}) we must collect all the terms which contribute coherently to
the sum.
Summing over these terms (weighted by ${\rm e}^{-\epsilon n}$) should provide a
pole at $\epsilon=0$, so that the final multiplication by $\epsilon $ will
yield
the residue at the pole. Inspecting (\ref {eq:firstsum}) we immediately
notice that,
for example,  the terms for which ${m'\over n+\nu}= {m\over n} = \alpha
(\nu) $ (where
$\alpha (\nu) $ depends only on $\nu$)   yield a contribution of the
desired nature,
because the net phase $\nu \Phi(2\pi \alpha (\nu) )$ is common to all the
summed
terms. To find these contributions in a consistent way, we shall identify
them as the
points where the first variation of the phase  of the summand vanishes.
\begin{eqnarray}
\delta \{ {\rm phase} \} & = &
\delta n \left [ \left(\Phi (2\pi {m'\over n+\nu })
-  \Phi (2\pi {m \over n  })\right ) \right . \nonumber \\
 && \ \ \ \ \  -  \left . 2\pi \left ( {m'\over n+\nu }\Phi'(2\pi
{m'\over n+\nu }) - {m \over n  }\Phi' (2\pi {m \over n  })\right ) \right
] \nonumber
\\ &+& \delta m' \left [2\pi \Phi'(2\pi {m'\over n+\nu }) \right ]\  - \
\delta m
\left [2\pi \Phi'(2\pi {m\over n })\right ]
\label{eq:stationary}
\end{eqnarray}
  The  phase is stationary with respect to variations in $n$  when
\begin{equation}
{m'\over n+\nu} = {m\over n}\ .
\label{eq:condition}
\end{equation}
One should consider only the solutions in the range  $m \le n $ and $m'\le
n+\nu $,
consistently with the range of the sums over $m$ and $m'$. The solution of
(\ref {eq:condition}) in integers is
\begin{eqnarray}
n&=&N \nu \ \ \ {\rm with} \ \ \ N=N_0,\cdots, \infty \nonumber \\
m&=&N k   \ \ \ {\rm with} \ \ \ k=\ 1,\cdots, \nu \nonumber \\
m'&=& (N+1) k \ .
\label {eq:integercond}
\end{eqnarray}
 The arbitrary integer $n_0$ is  replaced by another arbitrary constant
$N_0$ which
fixes the lower limit of the $n$ sum. This solution is the general solution for
the generic cases. One can always invent maps for which other stationary
points exist.
For each value of
$k$, the points of stationary phase form a grid. Near each grid point,  the
summation
variables will be replaced by  local variables
\begin{equation}
n= N\nu +\delta n  \ \ ; \ \ m= Nk +\delta m \ \ ; \ \ m'= (N+1)k +\delta
m' \ ,
\end{equation}
with
\begin{equation}
|\delta n| \le {\nu\over 2}  \ \ ; \ \  |\delta m| \le{ k\over 2} \ \ ; \ \
|\delta
m'|
\le{ k\over 2}
\ .
\end{equation}
The range of variation of the local variables is chosen such that each point in
the original sum will be counted once.   The contribution of the domain
about each
grid point will be computed by using the stationary phase approximation.
The  phase
cannot be made stationary  with respect to independent variations of  $m$
and $m'$
because there is no guarantee that $\Phi' (2\pi {m' \over n+\nu  })=0 $ and
$\Phi'
(2\pi {m\over n  })=0 $ have solutions when  $n$ , $m$ and $m'$ are
integers. However,
when (\ref {eq:condition}) is satisfied, the phase is stationary on the
manifold
$\delta m =\delta m'$.   Thus, the  sum over $n,m,m'$  in (\ref
{eq:firstsum}) is
replaced  by
\begin {equation}
\sum _{n=n_0}^{\infty}\ \sum_{m=1}^n\  \sum _{m'=1}^{n+\nu}\ \rightarrow \
\sum_{k=1}^{\nu}\  \sum _{N=N_0}^{\infty}\   \left \{ \sum _{\delta n =
-{\nu\over 2}}^{\nu\over 2}
  \ \sum_{\delta m =-{k\over 2}}^{k\over 2} \sum_{\delta m'
=-{k\over2}}^{k\over2}
{\delta } _{\delta m,\delta m'} \right \}
\end {equation}
Where the curly brackets on the  right enclose the contribution of the
vicinity of a
single grid point, restricted to the line $\delta m =\delta m'$.
On this line the summand is constant and therefore the
summation amounts to multiplication by  $ k$. The $\delta n$ sum can be
computed by considering the quadratic approximation to the phase near the
stationary
points and retaining the leading term in the result. It is determined by the
curvature of the phase at the point where it is  stationary
\begin{equation}
\left . {\partial^2 \{ {\rm phase} \} \over \partial n^2} \right |
=-(2\pi)^2 {k^2\over N(N+1) \nu^3} \Phi''(2\pi {k\over \nu}) \ .
\label{eq:curvature}
\end{equation}
The amplitude of the result depends on $N$, but when it is substituted in (\ref
{eq:firstsum}) it exactly cancels the $N$ dependent coefficient. The phase
of the
$\delta n$ sum is $\nu\Phi(2\pi {k\over\nu})$ which is also independent of $N$.
 Thus, the resulting terms of the $N$ sum depend on $N$ only through  ${\rm
e}^{-\epsilon \nu N}$.  Summing and taking the limit $\epsilon \rightarrow
0$ results
in a factor  $\nu ^{-1}$.  Collecting all the factors, one finds that (\ref
{eq:firstsum}) is  approximated by
\begin {eqnarray}
\label{eq:identity1scl}
&& \lim_{\epsilon \rightarrow 0} \ \epsilon \ \sum _{n=n_0}^{\infty} \left
[ t^{*}_n
 t_{n+\nu} \right ]_{\rm scl}{\rm
e}^{-n\epsilon}  =  \nonumber \\
&& \left ({2\pi \over \nu \hbar}\right )^{1\over2}
{\rm e}^ {- i( \nu+{1\over 2}) {\pi \over 2}}
\sum _{k=1}^{\nu}
\left [ \Phi''( 2\pi
{k\over \nu}) \right]^{1\over 2} {\rm e}^ { {i\over \hbar}  \nu\Phi(2\pi
{k\over\nu})}
\ \  = \
 \ \left [t_{\nu}\right ]_{\rm scl} \ .
\end{eqnarray}
 This completes the proof that the trace identities (\ref {eq:identity1}) are
compatible with the semiclassical approximation.

 The more complex relations (\ref {eq:identity2}) or the equivalent (\ref
{eq:identity3}) can be checked using the same technique.  As an example, the
compatibility of the identity
\begin{equation}
\lim_{\epsilon_1,\epsilon_2 =0} \epsilon_1 \epsilon_2 \sum_{n_1,n_2}
t^*_{n_1+n_2}t_{n_1+\nu_1}t_{n_2+\nu_2}{\rm e}^{-n_1\epsilon_1 -n_2\epsilon_2}
=t_{\nu_1+\nu_2}
\label {eq:threesum}
\end{equation}
will be demonstrated in some detail.

 Substituting in (\ref {eq:threesum}) the semiclassical approximation
(\ref{eq:trinteg}) for $t_n$, one has to consider the sum
\begin{eqnarray}
\label{eq:firstsum12}
 &\epsilon_1& \epsilon_2 {\rm e}^{-i(\nu_1+\nu_2) {\pi \over 2}}
\left ({2\pi \over\hbar}\right )^{3\over 2}
\sum_{n_1,n_2} {\rm e}^{-n_1\epsilon_1 -n_2\epsilon_2}
\left ({1\over n_1+n_2}{1\over n_1+\nu_1}{1\over n_2+\nu_2}\right )^{1\over2}
  \nonumber
\\ &\times &\sum_{m_1=1}^{n_1+\nu_1} \ \sum_{m_2=1}^{n_2+\nu_2}\
\sum_{m_{12}=1}^{n_1+n_2}
\left [\Phi'' (2\pi {m_1\over n_1+\nu_1}) \Phi'' (2\pi {m_2\over n_2+\nu_2})
\Phi'' (2\pi {m_{12}\over n_1+n_2}) \right ]^{1\over2}   \nonumber \\
&\times & {\rm e}^{{i\over \hbar}\left[ (n_1+\nu_1) \Phi (2\pi {m_1\over
n_1+\nu_1})
+(n_2+\nu_2)\Phi (2\pi {m_2\over n_2+\nu_2})
- (n_1+n_2) \Phi(2\pi {m_{12}\over n_1+n_2})\right ] }
\end{eqnarray}
 The phase of  (\ref {eq:firstsum12}) can be made stationary with respect to
variations of $n_1$ and $n_2$ under the following conditions:
\begin {eqnarray}
\label {eq:stat12}
-\Phi_{12} + \Phi_{1} + {m_{12}\over n_1+n_2}\Phi'_{12} -{m_{1}\over
n_1+\nu_1}\Phi'_{1} &=& 0 \\
-\Phi_{12} + \Phi_{2} + {m_{12}\over n_1+n_2}\Phi'_{12} -{m_{2}\over
n_2+\nu_2}\Phi'_{2} &=& 0
\nonumber
\end{eqnarray}
Where the short-hand notation $\Phi_1=\Phi (2\pi {m_1\over n_1+\nu_1})$ and
$\Phi'_1= 2\pi\Phi' (2\pi {m_1\over n_1+\nu_1})$, {\it etc} was used. The
conditions
(\ref {eq:stat12}) are satisfied  by  solutions in integers of:
\begin{equation}
{m_{12}\over n_1+n_2} = {m_{1}\over n_1+\nu_1} \le 1 \ \ {\rm and}\ \
{m_{12}\over
n_1+n_2} = {m_{2}\over n_2+\nu_2} \le 1 \ .
\label{eq:diophantine}
\end{equation}
The general solutions depend on three integers $N_1,\ N_2$ and $ k\ $ so that
\begin {eqnarray}
\label{eq:conditions12}
n_1 &=& (N_1-1)\nu_1 +N_1\nu_2   \ \ {\rm with } \ \  N_1=N_{0},\cdots, \infty
\nonumber \\
n_2 &=& N_2\nu_1 +(N_2-1)\nu_2   \ \ {\rm with } \ \  N_2=N_{0},\cdots, \infty
\nonumber \\ m_{12}& =& (N_1+N_2-1) k \ \ \ \ \ \ {\rm with} \ \  1\le k \le
\nu_1+\nu_2
\\ m_1 &=& N_1 \ k \nonumber \\
m_2 &=& N_2 \ k \nonumber
\end {eqnarray}
 Again, for each value of $k$ the points of stationary phase form a grid
and one
computes seperately the contribution from the vicinity of each grid point. The
summation volume about each grid point is of size
\begin {equation}
|\delta n_1, \delta n_2| \le {\nu_1 +\nu_2\over 2} \ \ ; \ \
|\delta m_1 ,\delta m_2 | \le {k\over 2} \ \ ; \ \  |\delta m_{12} | \le k
\label {eq:domain}
\end{equation}
   The phase cannot be made
stationary with respect to independent variations of $m_1,\ m_2 $ and $m_{12}$.
However, like in the previous case, the phase is constant for
\begin{equation}
\delta m_{12} = \delta m_1 +\delta  m_2
\label{eq:restrict}
\end{equation}

  The sums in (\ref {eq:firstsum12}) are rewritten in terms of the local
variables for each grid point, and the curly brackets encloses the sums over
individual domains (\ref {eq:domain}).
\begin{eqnarray}
& &\sum_{n_1=n_0}^{\infty}\sum_{n_2=n_0}^{\infty}\ \sum_{m_1  = 1}^{n_1
+\nu _1}\
\sum_{m_2=1}^{n_2 +\nu _2} \  \sum_{m_{12}=1}^{n_1+n_2}  \ \rightarrow  \\
& &
\sum_{k=1}^{\nu_1+\nu_2}\  \sum_{N_1=N_0}^{\infty}\sum_{N_2=N_0}^{\infty}\
\left \{
\sum_{\delta n_1}\  \sum_{\delta n_2}
\sum_{\delta m_1}\ \sum_{\delta m_2} \delta _{ \delta m_{12}, {
\delta} m_1+ \delta m_2} \right \} \nonumber
\end{eqnarray}
 The last Kronecker $\delta$ is due to the restriction (\ref
{eq:restrict}). Since the
summands in the  $\delta m_1, \delta m_2$ sums are constant, they
contribute a factor
of $k^2$. The $\delta n_1, \delta n_2$ sum is performed again by expanding
the phase to
second order, and retaining the leading term in the result. The determinant of
second derivatives at the point of stationary phase is
\begin{eqnarray}
&& \det \left ( {\partial^2 \{ {\rm phase} \} \over \partial n_i \partial n_j}
\right ) =  \\
&&-\left (2\pi{ k\over \nu _1+\nu _2}\right )^4
\left (\Phi''(2\pi{k\over \nu_1+\nu_2})\right )^2 {\nu _1+\nu _2\over
(n_1+n_2)(n_1+\nu _1)(n_2+\nu _2)} \ , \nonumber
\end{eqnarray}
where $n_1$ and $n_2$ take the values (\ref {eq:conditions12}).
Collecting all the factors, and performing the $N_1$ and $N_2 $ sums while
taking
the limit $\epsilon_1,\epsilon_2 \rightarrow 0$, one remains with the sum
over $k$
which can be easily identified as $\left [t_{\nu_1+\nu_2}\right ]_{\rm scl}$.

 This method  can be extended to identities involving higher powers. The
procedure becomes much more cumbersome and will not be reproduced here.

 The essence of the derivations outlined above is that the phase of the
summands is
constant on an infinite grid of integers. Only when  all of them are summed
together, they provide the necessary singularity which is cancelled against the
$\epsilon$ factors and give the correct answer.

The compatibility of the trace identities with the semiclassical approximation
demonstrate  the importance of ``non-diagonal" correlations which are
essentially due
to  the repetitive nature of the distribution of periodic orbits for integrable
maps.  As a matter of fact, had one applied the standard diagonal
approximation where
repetitions are neglected, to the sums  (\ref {eq:firstsum}),(\ref
{eq:firstsum12}) one
would get a vanishing result when $\nu \  {\rm or}  \ (\nu_1 +\nu_2) \ne 0$
. The
condition (\ref {eq:condition}) picks up (non-diagonal) pairs of
$n$-periodic and
$n+\nu$ periodic manifolds which coincide, since the action variable
$I_{n,m}$ is the
same.  This feature is effective in integrable systems since the
periodic manifolds are specified completely by integers (the period $n$ and
the winding
number $m$),and it is responsible for the compatibility of the
trace identity with the semiclassical approximation. The condition (\ref
{eq:conditions12}) expresses a similar coincidence of three periodic manifolds.
In chaotic systems, repetitions do exist but they play a much less
important r\^ole,
because apart from the period $n$ there exists no other integer which would
replace $m$  to specify the periodic orbits. This is why the methods
described in
the present section fails for the chaotic case.  The substitution of the
expressions
(\ref {trsemicl}) for systems which are chaotic in the classical limit, and
attempts to
identify the classical correlations which are implied by the trace
identities failed,
so far, to give a definite answer.

\section {\bf Discussion}
\label {sec:summary}
  In the previous section, the compatibility of the trace identities with the
semiclassical quantization of integrable maps was studied from the point of
view of
periodic orbit theory. It can be also proven very simply by observing that the
semiclassical expressions (\ref {eq:traceform}) for the $t_n$ has the formal
structure as a trace of a unitary matrix, and therefore
the trace identities which are based on nothing else but on this form, are
automatically ensured. This result does not detract from the work presented
in the
previous section, because it gave new insight into the interplay between
the trace
identities and the correlations in the spectrum of classical actions.

 In general, the semiclassical approximation for a quantum evolution
operator is the
leading approximant in an asymptotic expansion, in which the time is kept
fixed and
$\hbar$ approaches zero. Taking the opposite order in which $\hbar$ is
fixed and
the time is increased, is not allowed within the semiclassical
approximation. One has
to bear this basic rule  in mind when one tests the semiclassical
approximation. In
the present context, the magnitude of $\epsilon^{-1}$ sets the scale of the
relevant
evolution time (see (\ref {eq:timeaverage})). Hence the proper order of
limits would
seem to be such that  $\epsilon$ be fixed while $\hbar \rightarrow 0$.
However, the
semiclassical  limit  is synonymous to $M\rightarrow \infty$ (\ref {eq:
dimension}).
In this limit  the spectrum of $U$ becomes dense on the unit circle, and
the trace
identities do not hold. These conflicting demands can be satisfied  when
$\epsilon \le
{\cal O} ( \hbar ^{f})$. This remark does not apply to the test we
performed for
integrable maps, because the formal structure of  $[ t_n]_{\rm scl}$ as
given by (\ref
{eq:traceform})    is compatible with the trace identities for any values
of $\hbar$
and $n$. This is not the case for quantum maps which correspond to
classically chaotic
systems.

 So far, any attempt to assess the  compatibility of the trace identities with
the semiclassical approximation for classically chaotic systems ended in
failure.
Even  for maps on graphs,  where the periodic orbit expansion of $t_n$ is
exact \cite
{KS97,KS99,ScSm99}, I was unable to derive the trace identities within
periodic orbit
theory.  Therefore,  the nature of the underlying classical
correlations which ensure the exact trace identities when they are
expressed in terms
of periodic orbits remains an enigma, and calls for further research. In
Appendix A,
the trace identities are used to derive some combinatorial identities, which
illustrates an application outside of the domain of periodic orbits theory.

 \section {Acknowledgement}
 The author thanks the Humboldt foundation for the award which enabled him to
perform this research at the Philipps-Universitaet Marburg. The important
r\^ole of Professor Eugene Bogomolny to this work was mentioned in the text,
and it is a pleasure to thank him again. I am indebted to Dr Harel Primack
for his
critique and for the helpful comments and suggestions. Appendix A follows
from the
work on quantum graphs and combinatorics which is conducted in
collaboration with
Dr Holger Schanz, whose cooperation is acknowledged. I am indebted to
Professor  Bruno
Eckhardt for the hospitality in Marburg and for  useful discussions.
The work was  supported in part by the Minerva Center for Nonlinear Physics
of Complex
Systems, and by a grant from the Israel Science Foundation.

\appendix
\section{Identities for Krawtchouk polynomials}
We consider here the quantum map for a  $2$-star graph which was defined and
studied in detail in \cite {ScSm99}. It consists of a ``central" vertex
out of which
emerge $2$ bonds, terminating at vertices (with indices
$j=1,2$) with valencies $v_j=1$. The ratio between the bond lengths
 $L_j$ is assumed to be irrational. This simple model is not completely
trivial if the
central vertex scattering matrix is chosen as
\begin {equation}
\sigma^{(o)} =   \left (
{\begin{array} {ll}
 {1\over \sqrt {2}}     & {{\rm i}\over \sqrt {2}}   \\
 {{\rm i}\over \sqrt {2}}   & {1\over \sqrt {2}}
\end{array}}  \right )\,.
\label {2-starsigma}
\end{equation}
Neumann  boundary conditions are imposed at the two other vertices. The
Hilbert space
is of dimension $2$ and the evolution operator is
\begin {equation}
U(k) =
\left (
{\begin{array} {ll}
    {\rm e}^{2ikL_1} & 0   \\
  0    &  {\rm e}^{2ikL_2}
\end{array}}  \right )
\left (
{\begin{array} {ll}
 {1\over \sqrt {2}}     & {{\rm i}\over \sqrt {2}}   \\
 {{\rm i}\over \sqrt {2}}   & {1\over \sqrt {2}}
\end{array}}  \right )\,.
\label {2-starSB}
\end{equation}
where the diagonal matrix on the left takes care of the free propagation on
the bonds and the reflections from the vertices $j=1,2$.

 One can write an {\it exact} expression for $t_n(k) = {\rm tr} U^n(k)$ in
terms of
periodic orbits on the $2$-star graph which is analogous to (\ref
{trsemicl}). There
are $2^n / n$  $n$-periodic orbits. However, their lengths can take only
$n+1$ distinct
values : $L(n,q) = 2(q L_1  +(n-q)  L_2 )$, with $ 0\le q\le n$. Thus, one can
write
\begin{equation}
 t_n(k) = \sum_{q=0}^n {\rm e}^{i L(n,q) k} A(n,q) \ ,
\label{eq:trace2star}
\end{equation}
where $A(n,q)$ is the coherent sum of the amplitudes contributed by the
isometric
orbits with length $L(n,q)$. It can be computed explicitly \cite {ScSm99}
in terms of
Krawtchouk Polynomials
\begin{equation}
A(n,q) =  {1\over\sqrt{2^{n}}} \left \{
\begin{array}{l}
  1 \ \ \ \ \ \  \ \ \ \ \ \ \ \ \ \ \ \ \ \ \ \ \ \ \ \ \ \ \ \ \ \ \ \ \ \ \
 {\rm for}\ \ \ q=0 \ {\rm or}\ n  \nonumber \\
  (-1)^{n+q} \sqrt{ {n\over q}{n\choose q}  } P_{n-1,n-q} (q)  \ \
\  {\rm for} \ \ \ 0<q<n \ . \\
\end{array}
\right .
\end{equation}
and the Krawtchouk polynomials are defined as in \cite{kp1,kp2} by
\begin{equation}\label{eq:krawtchouk}
P_{N,k} (x)= {N\choose k}^{-1/2}\sum_{\nu=0}^{k}
(-1)^{k-\nu}{x\choose \nu}{N-x\choose k-\nu}
\ \ {\rm for} \ 0\le k \le N \  .
 \end{equation}
Substituting (\ref {eq:trace2star}) in the trace identity (\ref
{eq:identity1}),
we get for any integers $\nu$ and $n_0$
\begin {eqnarray}
\lim_{\epsilon \rightarrow 0} \ \epsilon &\sum_{n=n_0}^{\infty}&
{\rm e}^{-n\epsilon} \sum _{q=0}^n \sum_{p=0}^{n+\nu}
{\rm e}^{2ik\left[(p-q)L_1 +(\nu-(p-q))L_2\right ]}
A(n+\nu ,p) A(n,q) \nonumber \\
 & =&\sum_{\kappa=0}^{\nu}  A(\nu,\kappa) {\rm e}^{2ik\left [\kappa L_1
+(\nu-\kappa)L_2\right ]} \ .
\label{eq:identitykr}
\end{eqnarray}
This is valid for any value of the wave number $k$. When the  lengths
$L_1$ and $L_2$ are incommensurate, the equality can hold only if the
coefficients of
the phase factors ${\rm e}^{2ik L(\nu,\kappa)}$ on both sides are equal. This
implies
\begin {equation}
\lim_{\epsilon \rightarrow 0} \ \epsilon \ \sum _{n=n_0}^{\infty}
{\rm e}^{-n\epsilon} \sum _{q=0}^n A(n+\nu , q+\kappa) A(n,q)   = A(\nu,\kappa)
\ ,
\label{eq:identitykrav}
\end{equation}
with  $0 \le \kappa \le \nu $. Restricting to  $\kappa $ to the interval
$1 \le \kappa \le \nu-1 $ the identity can be expressed in terms of
 Krawtchouk polynomials exclusively.
\begin {eqnarray}
\label{eq:identitykravf}
& &\sqrt{ {\nu \over
\kappa}{\nu
\choose \kappa}  }\ P_{\nu-1,\nu -\kappa} (\kappa) =
\lim_{\epsilon \rightarrow 0} \ \epsilon \ \sum _{n=n_0}^{\infty}
{{\rm e}^{-n\epsilon}\over 2^n} \times    \\
&\times&\sum _{q=0}^n \sqrt{{n(n+\nu)\over q(q+\kappa)}
{n\choose q}{n+\nu\choose q+\kappa}  }\
 P_{n-1,n-q} (q)P_{n+\nu -1,n+\nu-q-\kappa}(q+\kappa)   \nonumber
\ .
\end{eqnarray}

 I could not find identities of this kind in the standard books on Krawtchouk
polynomials.

\section*{References}

\bibliographystyle{unsrt}

\end{document}